	\documentclass[prd,aps,tightenlines,nofootinbib,preprint]{revtex4}

\usepackage{graphicx}
\usepackage{amsmath,amssymb}

\def\mr{\mathrm}
\def\mb{\boldsymbol}
\def\mc{\mathcal}
\def\nm{\nonumber}

\def\eps{\epsilon}
\def\pd{\partial}

\newcommand{\dif}[3]{\frac{\mr{d}^{#1} #2}{\mr{d} {#3}^{#1}}}

\begin{document}

\preprint{YITP-12-50}

\title{Resonant Signatures of Heavy Scalar Fields in the Cosmic Microwave Background}
\author{Ryo Saito$^{1}$} \author{Masahiro Nakashima$^{2,3}$} \author{Yu-ichi Takamizu$^{1}$} \author{Jun'ichi Yokoyama$^{3,4}$}
\affiliation{
$^1$Yukawa Institute for Theoretical Physics, Kyoto University, Kyoto 606-8502, Japan\\
$^2$Department of Physics, Graduate School of Science,  				
The University of Tokyo, Tokyo 113-0033, Japan\\
$^3$Research Center for the Early Universe,
Graduate School of Science, The University of Tokyo,Tokyo 113-0033, Japan\\
$^4$Kavli Institute for the Physics and Mathematics of the Universe, The University of Tokyo, Chiba 277-8568, Japan
}

\begin{abstract}
We investigate the possibility that a heavy scalar field, whose mass exceeds the Hubble scale during inflation, could leave non-negligible signatures in the Cosmic Microwave Background (CMB) temperature anisotropy power spectrum through the parametric resonance between its background oscillations and the inflaton fluctuations. By assuming the heavy scalar field couples with the inflaton derivatively, we show that the resonance can be efficient without spoiling the slow-roll inflation. The primordial power spectrum modulated by the resonance has a sharp peak at a specific scale and could be an origin of the anomalies observed in the angular power spectrum of the CMB. In some values of parameters, the modulated spectrum can fit  the observed data better than the simple power-law power spectrum, though the resultant improvement of the fit is not large enough and hence other observations such as non-Gaussianity are necessary to confirm that the CMB anomalies are originated from the resonant effect of the heavy scalar field. The resonant signatures can provide an opportunity to observe heavy degrees of freedom during inflation and improve our understanding of physics behind inflation.
\end{abstract}

\maketitle


\section{Introduction}\label{s:intro}
 Cosmic inflation is the standard paradigm that provides the initial conditions for structure formation and the anisotropies in cosmic microwave background (CMB) as well as the global properties of the spacetime  \cite{Starobinsky:1980te, Sato:1980yn, Guth:1980zm}. In the inflationary scenario,  the accelerated expansion stretches quantum fluctuations on microscopic scales to cosmological scales, providing the seed for macroscopic observables as the anisotropies in the CMB. The wavelengths of the fluctuations are extremely short in the earliest epoch. Thus the cosmological observations provide a window into short-distance physics which is beyond the reach of terrestrial experiments.

The cosmological fluctuations, which have been generated quantum mechanically, are statistical in nature. In the simplest single-field slow-roll inflation models, they are approximately Gaussian-distributed and their power spectrum is nearly scale-invariant. Then we usually characterize them over the observable range of scales in terms of a power-law-type power spectrum, which is parametrized simply by two or three parameters: the amplitude, the spectral index, and the running. Though the two- (or three-) parameter spectrum consistently explains the observed CMB anisotropies \cite{Komatsu:2010fb}, it could miss valuable information on physics behind inflation. Recent high-resolution CMB data already implies the presence of fine features in the primordial power spectrum. Indeed, several groups, including one of the present authors (JY), reported statistically significant discrepancy between the prediction from a power-law primordial power spectrum and the CMB data \cite{TocchiniValentini:2005ja, Nagata:2008tk}. Using non-parametric reconstruction methods, they have found large anomalies in the reconstructed power spectrum, which are  localized around wavenumbers $k \simeq 0.003~\mr{Mpc}^{-1}$ and $k \simeq 0.009~\mr{Mpc}^{-1}$. On the other hand, a number of effects that cause deviations from a power-law power spectrum have been investigated in literatures, including trans-Planckian effects \cite{Martin:2000xs, Danielsson:2002kx,Schalm:2004qk}, a burst of particle production \cite{Romano:2008rr, Barnaby:2009dd}, temporal violation of slow-roll approximation \cite{Leach:2000yw, Saito:2008em, Starobinsky:1992ts, Adams:2001vc, Kaloper:2003nv, Battefeld:2010rf}, turns in the inflationary trajectory \cite{Burgess:2002ub, Achucarro:2010da, Shiu:2011qw}, a sharp water field transition \cite{Abolhasani:2012px}, and a sudden change of sound velocity \cite{Nakashima:2010sa}. They modify evolution of the inflaton fluctuations in their own way and leave their characteristic signatures on the primordial power spectrum. Thus fine features in the power spectrum could contain rich information such as detailed structures of  inflaton Lagrangian or existence of other degrees of freedom.
  
  In this paper, we consider an effect of coherent oscillations of a heavy scalar field whose mass exceeds the Hubble scale, $m \gg H$. Heavy scalar fields are ubiquitous in models of inflation embedded in supergravity and string theory. They appear as moduli fields, Kaluza-Klein modes, the scalar supersymmetric partner of inflaton, or others. In usual treatment, the dynamics of a heavy scalar field is neglected assuming that it is stuck to their potential minima during inflation because its excitation decays quickly \cite{Yamaguchi:2005qm}. The instantaneously-excited oscillations, however, can be important when we measure the primordial power spectrum with high resolution. An impact of the excitation has already been discussed in Ref. \cite{Burgess:2002ub} for a hybrid-inflation model. More recently, Refs. \cite{Shiu:2011qw, Cespedes:2012hu, Gao:2012} have discussed that oscillations excited by a sharp turn in multi-dimensional potential leave a ringing signature in the primordial power spectrum. In both cases, the signatures arise because the evolution of the fluctuations becomes non-adiabatic \cite{Martin:2000xs} by a sudden energy transfer between the inflaton and a heavy scalar field through their couplings in the potential. In addition, it has been discussed in Ref. \cite{Chen:2011zf} that a ringing feature in the power spectrum is induced through the gravitational couplings without considering direct couplings between inflaton and a heavy scalar field.
  
  In this paper, instead, we point out that a resonant enhancement of the fluctuations efficiently occurs deep in the horizon, $k/a \sim m \gg H$, through derivative couplings with a heavy scalar field. The derivative couplings are allowed even if a shift symmetry is imposed to ensure the flatness of the inflaton potential. Then there is no reason why they are absent in the action from the effective-field-theory point of view \cite{Weinberg:2008hq, Khosravi:2012qg}. Though the derivative couplings are usually irrelevant at low energy scale, they can play an important role on the evolution of the inflaton fluctuations in the resonance epoch. This is because the inflaton fluctuations and the heavy scalar field rapidly oscillate in the resonance epoch. Hence, effects of the derivative couplings on the evolution of the inflaton fluctuations are large there while their effects on the background evolution is relatively small because the derivative of the background inflaton field is slow-roll suppressed. In the following sections, we estimate the enhancement of the primordial power spectrum by the resonance assuming non-derivative couplings are sufficiently suppressed. In contrast to other effects as the slow-roll violation \cite{Kumazaki:2011eb}, the feature induced by the resonance can be sharp and large even in the case that adiabaticity is mildly violated because the resonance coherently accumulates small effects. The instantaneously excited oscillations do not affect the slow-roll background evolution much in this case unless the heavy scalar field dominates the energy density because the flatness of the inflaton potential is ensured even during the oscillations.
  
  The organization of this paper is as follows. In \S \ref{s:model}, we present our model to realize an efficient enhancement of the fluctuations and discuss conditions on the model required for the slow-roll inflation. In \S \ref{s:amp}, we estimate the enhancement of the fluctuations by the parametric resonance and discuss consistency of our model with the anomalies observed in the CMB spectrum. Finally, we provide our summary of this paper in \S \ref{s:summary}. 
  

\section{The model}\label{s:model}
 In this section, we introduce our model and discuss conditions on model parameters required for the slow-roll inflation.


\subsection{An inflationary model with a heavy scalar field}
 We consider a model with a heavy scalar field with mass $m \gg H$ which couples to the inflaton field in the following manner,
	\begin{align}
		S_{m} &\equiv -\int \mr{d}x^4 \sqrt{-g}\left[ \frac{1}{2}(\pd \phi)^2 + V(\phi) + \frac{1}{2}(\pd \chi)^2 + \frac{m^2}{2}\chi^2 + K_n + K_d \right], \label{eq:action}
	\end{align}
with
	\begin{align}
		K_n  &\equiv \frac{\lambda_n}{2\Lambda_n}\chi(\pd \phi)^2, \label{eq:ncouple}
	\end{align}
and
	\begin{align}
		K_d  &\equiv \frac{\lambda_{d1}}{4\Lambda_d^4}(\pd \chi)^2(\pd \phi)^2 + \frac{\lambda_{d2}}{4\Lambda_d^4}(\pd \chi \cdot \pd \phi)^2, \label{eq:dcouple}
	\end{align}
where $\phi$ is the inflaton field and $\chi$ is the heavy scalar field. The potential $V(\phi)$ is assumed to be sufficiently flat;
	\begin{align}\label{eq:slowroll}
		\eps_V \ll 1,~|\eta_V| \ll 1,
	\end{align}
where
	\begin{equation}
		\eps_V \equiv \frac{M_p^2}{2}\left(\frac{V'}{V}\right)^2, \quad \eta_V \equiv M_p^2\frac{V''}{V},
	\end{equation}
are the slow-roll parameters. Here, $M_p=2.4 \times 10^{18}\mr{GeV}$ is the reduced Planck mass.

The heavy scalar field $\chi$ is assumed to be subdominant;
	\begin{align}\label{eq:sub}
		f_{\chi} \ll 1,
	\end{align}
where
	\begin{equation}
		f_{\chi} \equiv \frac{\rho_{\chi}}{\rho} \simeq \frac{\dot{\chi}^2+m^2\chi^2}{6M_p^2H^2},
	\end{equation}
is the fraction of its energy density to the total one. We assume that $\chi$ decays with a rate $\Gamma$, which satisfies $H \ll \Gamma \ll m$.

 The derivative couplings $K_n$ and $K_d$ are expected to appear in the action if we consider the action (\ref{eq:action}) as the leading terms in a generic effective field theory \cite{Weinberg:2008hq, Khosravi:2012qg}. Though other terms are also allowed in general, the derivative couplings $K_n$ and $K_d$ provide the most general couplings between the inflaton and the heavy scalar field at the leading order in $1/\Lambda_n$ and $1/\Lambda_d$ in a model with a parity symmetry, $\phi \to -\phi$, and a shift symmetry, $\phi \to \phi+c$. Moreover, the couplings $K_d$ cannot be forbidden while the couplings $K_n$ can be suppressed by imposing a shift symmetry on $\chi$. Higher-order terms in $1/\Lambda_n$ and $1/\Lambda_d$ are also expected to appear in general, though we have suppressed them in Eq. (\ref{eq:action}). To ensure that contributions from these terms can be safely neglected, we assume hereafter that the background fields satisfy the following conditions, 
	\begin{equation}\label{eq:irrelevant}
		\chi \ll \Lambda_n, \quad \dot{\phi},~\dot{\chi} \ll \Lambda_d^2.
	\end{equation}
We have also suppressed terms $(\pd \phi)^4$ and $(\pd \chi)^4$, because they have little effect on our analysis under the conditions above. In general, non-derivative couplings as $\chi V(\phi)$ could appear since the potential breaks the shift symmetry. However, we assume the non-derivative couplings are sufficiently suppressed in the following analysis for simplicity. Even if we include non-derivative couplings, they modify only the background evolution and do not directly affect the evolution of fluctuations much.

 We can also find the couplings $K_n$ and $K_d$ in specific models of inflation. For example, the couplings $K_n$ naturally appear if the inflaton is a pseudo-Nambu Goldstone boson, where the heavy scalar field is provided by the symmetry breaking field. Moreover, they can be found in the Einstein-frame action if a model contains additional scalar degrees of freedom with a non-minimal coupling to curvature, and also in the supergravity action with higher-order terms in the K\"{a}hler potential. In the latter case, the scalar supersymmetric partner of inflaton can be a candidate for the heavy scalar field, which usually has large mass corrections from the Planck suppressed terms in the F-term potential since it is not protected by a shift symmetry. On the other hand, the couplings $K_d$ can be found in models such as brane inflation. In brane inflation, for example, we can see that they appear in the DBI action by expanding its square root with the heavy scalar field being provided by KK modes or brane coordinates other than the inflaton.
 In the case of brane inflation models, where all the higher-order terms neglected in the action (\ref{eq:action}) are known, we may not need to impose the conditions (\ref{eq:irrelevant}) though instead the evolution equation should be solved including their effects. In addition, they have non-derivative couplings and hence we should make the analysis including them. These cases will be discussed in another paper.


\subsection{The background evolution}
 The evolution equations for the background fields can be written as
	\begin{align}
		\dot{\pi}_{\phi} + 3H\pi_{\phi} + V' = 0, \label{eq:ibgeom} \\
		\dot{\pi}_{\chi} + (3H+\Gamma)\pi_{\chi} + m^2\chi + \frac{\lambda_n}{2\Lambda_n}\dot{\phi}^2= 0, \label{eq:hbgeom}
	\end{align}
where $\pi_{\phi}$ and $\pi_{\chi}$ are conjugate momenta of the scalar fields,
	\begin{align}
		\pi_{\phi} &\equiv \left[ 1 + \lambda_n\frac{\chi}{\Lambda_n} + \left(\lambda_{d1}+\lambda_{d2}\right)\frac{\dot{\chi}^2}{2\Lambda_d^4} \right]\dot{\phi} \label{eq:icm} \\
		&\equiv K_{1I}\dot{\mb{\phi}}^{I},\\
		\pi_{\chi} &\equiv \left[ 1 + \left(\lambda_{d1}+\lambda_{d2}\right)\frac{\dot{\phi}^2}{2\Lambda_d^4} \right]\dot{\chi} \label{eq:hcm}\\
		&\equiv K_{2I}\dot{\mb{\phi}}^{I}.
	\end{align}
 Here we have introduced a notation that
	\begin{equation}
		K \equiv \frac{1}{2}(\pd \phi)^2 + \frac{1}{2}(\pd \chi)^2 + K_n + K_d,
	\end{equation}
and $\mb{\phi}^{(1)} \equiv \phi,~\mb{\phi}^{(2)} \equiv \chi$ for brevity.\footnote{We use the summation convention for repeated indices throughout the paper and parenthesis for components in the field space.} $K_{IJ}$ represents $K$ differentiated by $X^{IJ} \equiv -\pd \mb{\phi}^I \cdot \pd \mb{\phi}^J/2$. A dissipation term, $\Gamma \pi_{\chi}$, has been added in Eq. (\ref{eq:hbgeom}) to incorporate the effect of the decay of $\chi$.

 We can obtain the solutions for the background evolution as usual,
	\begin{align}
		\pi_{\phi}(t) &\simeq -\frac{V'}{3H} \qquad (\text{slow-roll solution}), \label{eq:ibgsol} \\
		\chi(t) &\simeq \chi_0 e^{-\Gamma t}\cos(mt), \label{eq:hbgsol}
	\end{align}
if the conditions (\ref{eq:irrelevant}) are satisfied at the onset of the oscillations, $t=0$. We discuss consistency of these solutions in some details here.

 Before proceeding to a discussion on the consistency of the solutions (\ref{eq:ibgsol}) and (\ref{eq:hbgsol}), we discuss whether inflation occurs or not during the oscillations. Inflation is realized if the slow-variation parameter is small; 
	 \begin{align}\label{eq:eps}
		\eps_H \equiv -\frac{\dot{H}}{H^2} \ll 1.
	\end{align}
 Using the Friedman equation, $\dot{H}=-K_{IJ}X^{IJ}/M_p^2$, the equation (\ref{eq:eps}) can be rewritten as
	\begin{align}
		\eps_H &= \frac{K_{IJ}X^{IJ}}{M_p^2H^2} \nm \\
		&\simeq \frac{1}{2}\left(\frac{\dot{\phi}}{M_p H}\right)^2 + \frac{1}{2}\left(\frac{\dot{\chi}}{M_p H}\right)^2, \label{eq:eps2}
	\end{align}
where we have neglected the higher-order terms in $1/\Lambda_n$ and $1/\Lambda_d$. Again, using the Friedman equation, $3M_p^2 H^2 = \rho \simeq V$, and the solution (\ref{eq:hbgsol}), we can further rewrite Eq. (\ref{eq:eps2}) as 
	\begin{align}
		\eps_H &\simeq \eps_V + \frac{3}{2}f_{\chi}\sin^2(mt), \label{eq:eps3}
	\end{align}
where we have neglected the higher-order terms in $\eps_V$ and $f_{\chi}$. Hence, the conditions (\ref{eq:slowroll}), (\ref{eq:sub}), and (\ref{eq:irrelevant}) are sufficient for realizing inflation. Note that another slow-variation parameter $\eta_H \equiv \dot{\eps}_H/H\eps_H$ is not so small  during the oscillations due to the second term. 

 In order that the slow-roll solution (\ref{eq:ibgsol}) be consistent, the first term should be smaller than the others in Eq. (\ref{eq:ibgeom}); $\dot{\pi}_{\phi}/3H\pi_{\phi} \ll 1$. Substituting the solution (\ref{eq:ibgsol}), we obtain
	\begin{align}
		\frac{\dot{\pi}_{\phi}}{3H\pi_{\phi}} &\simeq \frac{V''}{V'}\frac{\dot{\phi}}{H} - \frac{\dot{H}}{H^2} \nm \\
		&\simeq -\eta_V +\eps_H, \label{eq:iconsistency}
	\end{align}
which shows that the solution (\ref{eq:ibgsol}) can be used consistently under the conditions (\ref{eq:slowroll}), (\ref{eq:sub}), and (\ref{eq:irrelevant}). It is noteworthy that we cannot replace $\dot{\pi}_{\phi}$ by $\ddot{\phi}$ in deriving Eq. (\ref{eq:iconsistency}). The differentiation of $\chi$ in Eq. (\ref{eq:icm}) induces a large factor $m$, hence $\ddot{\phi}/3H\dot{\phi}$ can be large,
	\begin{align}
		\frac{\ddot{\phi}}{3H\dot{\phi}} &\sim O\left(\frac{m}{H}\frac{\chi}{\Lambda_n}\right) + O\left(\frac{m}{H}\frac{\dot{\chi}^2}{\Lambda_d^4}\right).
	\end{align}
 Finally, $\dot{\pi}_{\chi}$ can be replaced by $\ddot{\chi}$ in solving Eq. (\ref{eq:hbgeom}) because both $\dot{\phi}$ and $\dot{\chi}$ induce the factor $m$. Hence the couplings $K_d$ do not affect the solution (\ref{eq:hbgsol}). On the other hand, the last term in Eq. (\ref{eq:hbgeom}) induces an approximately constant term in $\chi$ as
	 \begin{align}\label{eq:chishift}
		\chi(t) \simeq -\frac{\lambda_n \dot{\phi}^2}{2m^2\Lambda_n} + \left( \chi_0+\frac{\lambda_n \dot{\phi}^2}{2m^2\Lambda_n} \right) e^{-\Gamma t}\cos(mt).
	\end{align}
Though the constant term does not spoil the resonance, it should satisfy the condition (\ref{eq:irrelevant}). This condition can be satisfied if $m$ is sufficiently large,
	\begin{equation}\label{eq:nccond}
		\frac{m}{H} \gg \sqrt{\lambda_n\eps_{H,\phi}} \left(\frac{\Lambda_n}{M_p}\right)^{-1},
	\end{equation}
where $\eps_{H,\phi}$ is the contribution of the inflaton to $\eps_H$. The constant term also contributes to the potential energy. However, the induced potential energy is always subdominant if the condition (\ref{eq:nccond}) is satisfied. In the case that the constant term is comparable to the initial amplitude $\chi_0$, the amplitude of the oscillations is provided as Eq.(\ref{eq:chishift}) instead of $\chi_0$.

 In summary, the oscillations of the heavy scalar field $\chi$ do not spoil the slow-roll inflation provided that the conditions (\ref{eq:slowroll}), (\ref{eq:sub}), and (\ref{eq:irrelevant}) are satisfied. In the next section, we investigate the evolution of the inflaton fluctuations in this background and show that a parametric amplification of these fluctuations takes place through the resonance with the oscillations of $\chi$.
 

\section{Parametric resonance with the oscillations of the heavy scalar field}\label{s:amp}
 In the previous section, we saw that the oscillations of the heavy scalar field do not much affect the background evolution provided that the conditions (\ref{eq:slowroll}), (\ref{eq:sub}), and (\ref{eq:irrelevant}) are satisfied. In this section, we show that the fluctuations in the inflaton field can be enhanced through the parametric resonance with the oscillations of $\chi$ even in this case.\\


\subsection{Evolution equation for the inflaton fluctuations}
 First, we derive an evolution equation for the inflaton fluctuations. The heavy scalar field $\chi$ oscillates with a frequency $m \gg H$, then the resonance occurs on scales much smaller than the horizon scale. Hence, we can neglect contributions from the metric fluctuations during the resonance because they are much smaller than those in the scalar fields on subhorizon scales as explicitly shown in Appendix \ref{a:assumptions}. Furthermore, the fluctuations in the heavy scalar field can be neglected because they do not contribute to the resonance. To eliminate the gauge degrees of freedom in the metric fluctuations, we employ the flat gauge, where the spatial metric becomes $a^2\delta_{ij}$.
 
  Neglecting the contributions from the fluctuations in the metric and the heavy scalar field, the second-order action in the inflaton fluctuations, $\varphi$, can be written as,
	\begin{align}
		S_2 &\simeq \int \mr{d}t\mr{d}^3x~ \frac{a^3}{2}\left[ (2K_{1K,L1}X^{KL}+K_{11})\dot{\varphi}^2 - K_{11}(\nabla \varphi)^2/a^2 - K_{1,1}\varphi^2 \right] \\
		&\simeq \int \mr{d}t\mr{d}^3x~ \frac{z_{\phi}^2}{2}\left[ \dot{\varphi}^2 - c_s^2(\nabla \varphi)^2/a^2 \right], \label{eq:2action}
	\end{align}
where
	\begin{align}
		z_{\phi}^2 &\equiv a^3(2K_{1K,L1}X^{KL}+K_{11})\\
		&= a^3\left[1 + \lambda_n\frac{\chi}{\Lambda_n} + \left(\lambda_{d1}+2\lambda_{d2}\right)\frac{\dot{\chi}^2}{2\Lambda_d^4}\right], \\
		c_s^2 &\equiv \frac{K_{11}}{2K_{1K,L1}X^{KL}+K_{11}} \\
		&\simeq 1 + \left(\lambda_{d1}-2\lambda_{d2}\right)\frac{\dot{\chi}^2}{2\Lambda_d^4} + O\left(\frac{\dot{\chi}^4}{\Lambda_d^8}\right).
	\end{align}
 Here, we have neglected the potential term, $K_{1,1}\varphi^2 \simeq -3\eta_V H^2 \varphi^2$, which is much smaller than the term $(\nabla \varphi)^2/a^2$ during the resonance. Note that non-derivative couplings, if any, can be similarly neglected unless the slow-roll conditions (\ref{eq:slowroll}) are violated. The action (\ref{eq:2action}) leads to the following evolution equation in Fourier space for the inflaton fluctuations,
	\begin{equation}\label{eq:emfl}
		\ddot{v}_{k} + \left[c_s^2\left(\frac{k}{a}\right)^2 - \frac{\ddot{z}_{\phi}}{z_{\phi}}\right]v_{k}=0,
	\end{equation}
where $v \equiv z_{\phi}\varphi$. Neglecting the higher-order terms in $\chi/\Lambda_n$ and $\dot{\chi}/\Lambda_d^2$, the quantities $c_s^2$ and $\ddot{z}_{\phi}/z_{\phi}$ are estimated to be
	\begin{align}
		c_s^2 &\simeq 1 + (\lambda_{d1}-2\lambda_{d2})\frac{m^2\hat{\chi}_0^2}{2\Lambda_d^4}\sin^2(mt), \label{eq:sonic}\\
		\begin{split}
		\frac{\ddot{z}_{\phi}}{z_{\phi}} &\simeq \frac{3}{4}(3-2\eps_H)H^2 + 3mH\left[\lambda_n\frac{\hat{\chi}_0}{\Lambda_n}\sin(mt) + (\lambda_{d1}+2\lambda_{d2})\frac{m^2\hat{\chi}_0^2}{2\Lambda_d^4}\sin(2mt)\right] \\
		& \hspace*{.25\linewidth} + m^2\left[\lambda_n\frac{\hat{\chi}_0}{2\Lambda_n}\cos(mt) + (\lambda_{d1}+2\lambda_{d2})\frac{m^2\hat{\chi}_0^2}{2\Lambda_d^4}\cos(2mt)\right] \nm
		\end{split}\\
		&\simeq m^2\left[\lambda_n\frac{\hat{\chi}_0}{2\Lambda_n}\cos(mt) + (\lambda_{d1}+2\lambda_{d2})\frac{m^2\hat{\chi}_0^2}{2\Lambda_d^4}\cos(2mt) + O\left(\frac{H}{m}\right)\right], \label{eq:efmass}
	\end{align}
where $\hat{\chi}_0 \equiv \chi_0e^{-\Gamma t}$. Substituting Eq. (\ref{eq:sonic}) and Eq. (\ref{eq:efmass}) into Eq. (\ref{eq:emfl}), the evolution equation can be rewritten in the form,
	\begin{equation}\label{eq:hill}
		\ddot{v}_{k} + m^2\left[\left(\frac{k}{am}\right)^2 - \frac{q_n}{2}\cos(mt) - 2q_d\cos(2mt) \right]v_{k}=0,
	\end{equation}
where
	\begin{align}
		q_n &\equiv \lambda_n\frac{\hat{\chi}_0}{\Lambda_n}, \label{eq:qn} \\
		q_d &\equiv -(\lambda_{d1}-2\lambda_{d2})\frac{m^2\hat{\chi}_0^2}{8\Lambda_d^4}\left(\frac{k}{am}\right)^2 + (\lambda_{d1}+2\lambda_{d2})\frac{m^2\hat{\chi}_0^2}{4\Lambda_d^4}. \label{eq:qd}
	\end{align}
 Hereafter, we consider the limit where either $q_n$ or $q_d$ is much larger than the other for simplicity. In this case, the evolution equation (\ref{eq:hill}) can be written in the form of the Mathieu equation,
	\begin{equation}\label{eq:mathieueq}
		\dif{2}{{v}_{k}}{z} + \left[A_k - 2q\cos(2z) \right]v_{k}=0,
	\end{equation}
where $(z,A_k,q) \equiv (mt/2,(2k/am)^2,q_n)$ for the $K_n$-dominated case and $(z,A_k,q) \equiv (mt,(k/am)^2,q_d)$ for the $K_d$-dominated case. Here, the value of $q$ is smaller than unity because $\chi/\Lambda_n \ll 1$ and $m\chi/\Lambda_d^2 \sim \dot{\chi}/\Lambda_{d}^2 \ll 1$ as assumed in the previous section, and then the adiabaticity is mildly violated. Hence the resonance is narrow, which occurs in  a narrow instability band, $|A_k-1|<q$. The parametric resonance occurs when the modes are redshifted to this instability band; 
	\begin{align}
		\left(1-\frac{q_n}{2}\right)\frac{m}{2} < \frac{k}{a} < \left(1+\frac{q_n}{2}\right)\frac{m}{2},
	\end{align}
for the $K_n$-dominated case and
	\begin{align}
		\left(1-\frac{\tilde{q}_d}{2}\right)m < \frac{k}{a} < \left(1+\frac{\tilde{q}_d}{2}\right)m,
	\end{align}
for the $K_d$-dominated case where
	\begin{equation}\label{eq:qdtilde}
		\tilde{q}_d \equiv (\lambda_{d1}+6\lambda_{d2})\frac{m^2\hat{\chi}_0^2}{8\Lambda_d^4}.
	\end{equation}
 In both cases, the resonance occurs in a deep subhorizon regime since the mass scale $m$ is much larger than the Hubble scale, $m \gg H$. Though we consider only the case that one dominates the other, the resonance is expected to be efficient even when both of the couplings are non-negligible since the values of their oscillation frequencies are rationally related \cite{Braden:2010wd}.

 After the oscillations damp out, we can use the standard evolution equation for the inflaton field in a single-field inflation model, which can be written as
	\begin{equation}\label{eq:stemlf}
		\ddot{v}_{k} + \left[\left(\frac{k}{a}\right)^2 - \hat{\mc{M}}H^2 \right]v_{k}=0,
	\end{equation}
where
	\begin{align}
		\hat{\mc{M}} &\equiv \frac{3}{4}(3-2\eps_H) + \frac{3}{2}\eta_H - \frac{1}{2}\eps_H\eta_H + \frac{1}{4}\eta_H^2 + \frac{\dot{\eta}_H}{2H} \\
		&= \frac{3}{4}(3-2\eps_H) + \frac{1}{a^3 M_p^2 H^2}\dif{}{}{t}\left(\frac{a^3\pi_{\phi}^2}{H}\right) - \frac{V''}{H^2},
	\end{align}
in our notation. Here, we have added the contributions from the metric fluctuations and the potential term, which induce the global tilt of the spectrum.
 
	\begin{figure}[h]
		\centering
		\includegraphics[width=.6\linewidth]{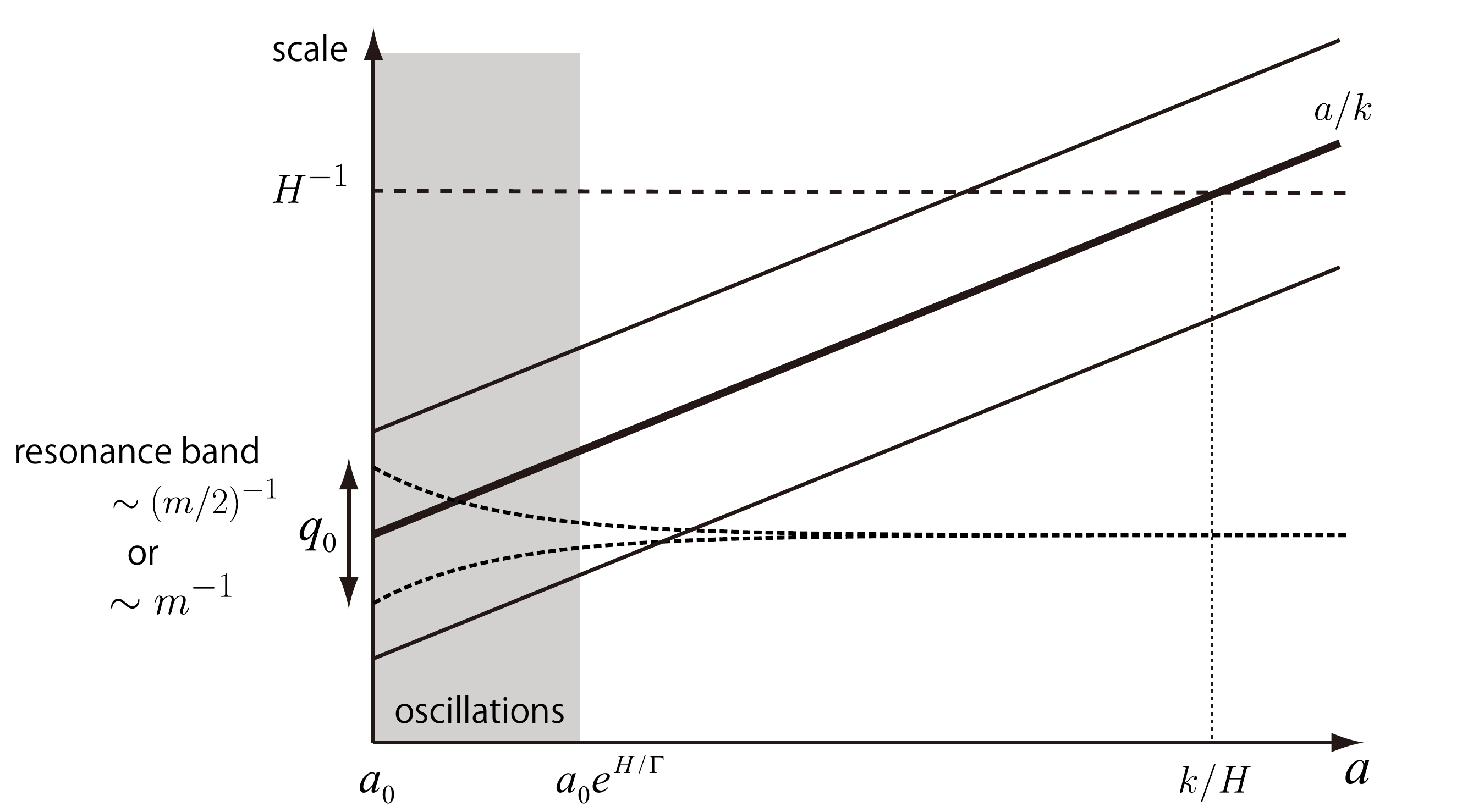}
		\caption{Time evolution of the inflaton fluctuations. The fluctuations on observable scales can be enhanced through the parametric resonance when they are deep in the horizon. The modes with comoving scales represented by the thick line are enhanced by the resonance, while those represented by the thin lines are not because they never cross the resonance band during the oscillations of the heavy scalar field. Hence, the resonance induces a peak in the power spectrum. Here, the variables with subscript $0$ indicate those evaluated at the onset of the oscillations.}
		\label{fig:te}
	\end{figure}

	
 \subsection{Parametric amplification of the curvature perturbations}
 We estimate the amplification of the power spectrum for the comoving curvature perturbations relative to the standard one. The comoving curvature perturbations, $\zeta$, are proportional to the inflaton fluctuations in the flat gauge, $\varphi$ at linear order as $\zeta=H\varphi/\dot{\phi}$ after the oscillations damp out. Hence, the ratio of the modulated power spectrum to the unmodulated one, $\mc{A}$, can be written by the mode function for the inflaton fluctuations as,
	\begin{equation}
		\mc{A} = \lim_{t \to \infty} \left|\frac{\widetilde{\varphi}_k}{\varphi_k}\right|^2=\lim_{t \to \infty} \left|\frac{\widetilde{v}_k}{v_k}\right|^2,
	\end{equation}
where we have denoted the modulated quantity by a symbol with tilde.  We present the results for the $K_d$-dominated case here. The analysis for the $K_n$-dominated case can be made in a  similar way. 

The spectral shape of the modulated power spectrum can be roughly understood from Fig. \ref{fig:te}. The spectrum has a peak around the modes which had the mass scale $\sim m$ at the onset of the oscillations and the peak width for smaller and larger wavenumbers, $\Delta_S$ and $\Delta_L$, are respectively determined by the width of the resonance band $q_0$ and the duration of the oscillations $H/\Gamma$. From Fig. \ref{fig:te}, we can see that the modes whose wavenumbers are around $k_p/a_0 \equiv m(1+q_0/2)$ are most amplified while the modes whose wavenumbers are less than $k_S/a_0 \equiv m(1-q_0/2)$ are hardly amplified since they were already outside of the resonance band at the onset of the oscillations. Hence, the peak width for smaller wavenumbers is roughly given by,
	\begin{align}
		\Delta_S \simeq \left| \frac{k_p}{a_0 m} - \frac{k_S}{a_0 m} \right| \sim q_0.
	\end{align}
 On the other hand, the modes whose wavenumbers are larger than $k_L/a_0 \equiv e^{H/\Gamma}m$ are also hardly amplified because they cross the resonance band after the oscillations damped out. Hence, the peak width for larger wavenumbers is roughly given by,
	\begin{align}
		\Delta_L \simeq \left| \frac{k_p}{a_0 m} - \frac{k_L}{a_0 m} \right| \sim \frac{H}{\Gamma}.
	\end{align}
 Moreover, the peak amplitude can be roughly estimated by using the Floquet exponent \cite{Allahverdi:2010xz}, $\mu=q/2$, as,	 
	 \begin{align}
		v_{k} \propto \exp\left(\int \mu {\rm d}z \right),
	\end{align}
 where the exponent is approximately given by,
 	\begin{align}
		\int \mu {\rm d}z &\simeq 
		\begin{cases}
		{\displaystyle \frac{mq_0}{4\Gamma}} & \text{for} \quad q_0 \gg H/\Gamma, \\
		{\displaystyle \frac{mq_0^2}{4H}} & \text{for} \quad q_0 \ll H/\Gamma,
		\end{cases}\\
		&\sim {\displaystyle \frac{mq_0}{4H}\min\left(q_0, \frac{H}{\Gamma}\right)}. \label{eq:amp}
	\end{align}
 Here, the integration range is determined by the condition $|A_k-1|<q$. 
 
\begin{figure}[phtb]
		\centering
		\begin{minipage}{.45\linewidth}
		\includegraphics[width=.95\linewidth]{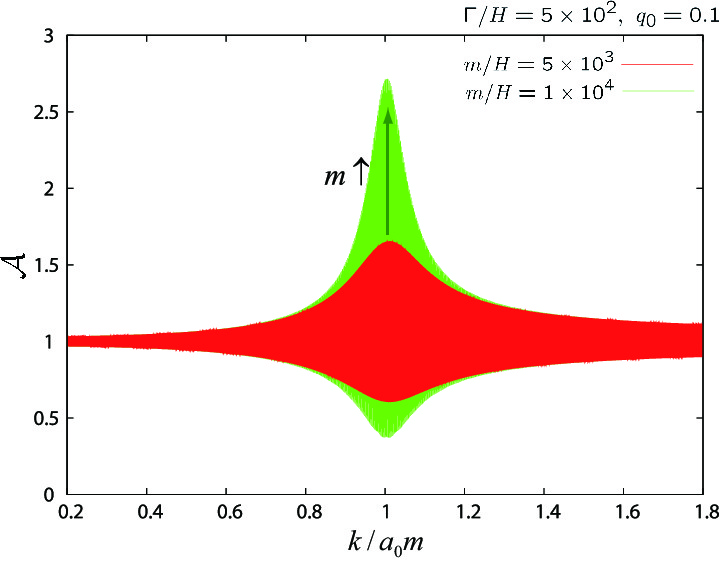}
		\end{minipage}
		\begin{minipage}{.45\linewidth}
		\includegraphics[width=.95\linewidth]{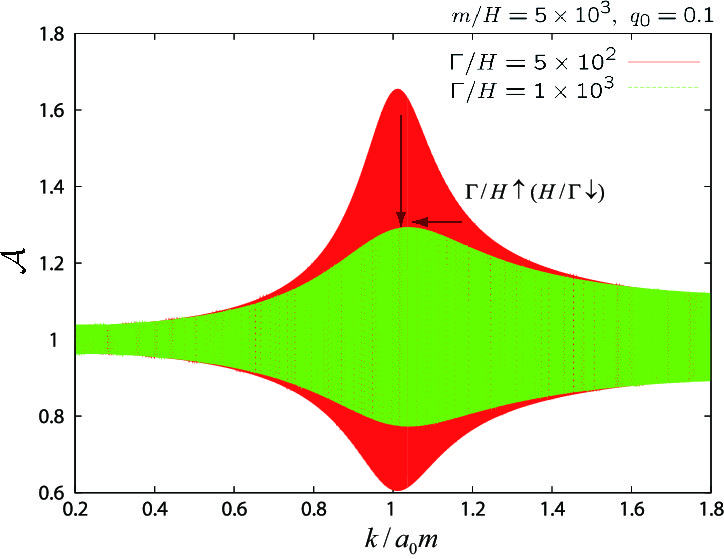}
		\end{minipage}
		\begin{minipage}{.45\linewidth}
		\includegraphics[width=.95\linewidth]{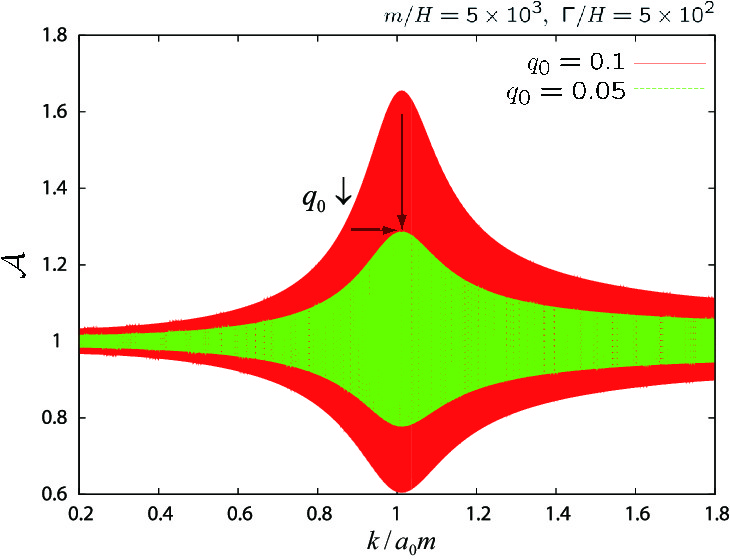}
		\end{minipage}
		\caption{The amplification of the power spectrum for different values of the parameters, $m$ (top left panel), $\Gamma$ (top right panel), and $q_0$ (bottom panel) for the $K_d$-dominated case. The spectra behave similarly in the $K_n$-dominated case.}
		\label{fig:primordial}
	\end{figure}

 In Fig. \ref{fig:primordial}, we have shown the amplification of the power spectrum, $\mc{A}$, estimated numerically assuming that the inflaton fluctuations were in the Bunch Davis vacuum before the oscillations start. Each panel shows the spectra for different values of a parameter while the others are fixed. The resultant spectra exhibit the expected dependence on the parameters and rapidly oscillate with frequencies of the order of $m/H$.  In the $K_n$-dominated case, the spectrum behaves similarly.
 
 Since the width and  the amplitude depend on the model parameters in different ways, we can extract information on the heavy scalar field from the features in the primordial power spectrum. For example, we can read values of the model parameters in the $K_d$-dominated case from the data as,
 	\begin{subequations}\label{eq:modelparam}
	 \begin{align}
		\Gamma &\sim 10^{13}~{\rm GeV}\left(\frac{H}{10^{12}~{\rm GeV}}\right)\left(\frac{\Delta_L}{0.1}\right)^{-1}, \\
		m &\sim 10^{14}~{\rm GeV}\left(\frac{H}{10^{12}~{\rm GeV}}\right)\left(\frac{\ln \mathcal{A}_p}{1.0}\right)\left(\frac{\Delta_S}{0.1}\right)^{-1}\left(\frac{\Delta_{\rm min}}{0.1}\right)^{-1}, \\
		\Lambda_d &\sim 10^{15}~{\rm GeV}\left(\frac{H}{10^{12}~{\rm GeV}}\right)^{1/2}\left(\frac{f_{\chi}}{0.1}\right)^{1/4}\left(\frac{\Delta_S}{0.1}\right)^{-1/4},
	\end{align}
	\end{subequations}
assuming $\lambda_{d1}$ and $\lambda_{d2}$ being ${\cal O}(1)$. Here, we have made a rough estimation of the width as $\Delta_S \sim q_0$ and $\Delta_L \sim H/\Gamma$ and the amplification at the peak scale ${\cal A}_p$ has been estimated as Eq. (\ref{eq:amp}). We have also introduced $\Delta_{\rm min} \equiv \min(\Delta_S, \Delta_L)$ for brevity.
 
 
 \subsection{Comparison with observational data of CMB}
 Having clarified the dependence of the modulated power spectrum on our model parameters, we can in principle fix or constrain their values by confronting them with observational data. In practice, however, it is difficult to find best-fit parameters for this kind of highly oscillatory spectrum because there are many minima in the likelihood surface and then the Markov Chain Monte Carlo chains do not converge \cite{Flauger:2009ab, Meerburg:2011gd}. Moreover, the parameter search is time-consuming because we do not have the exact analytic expression of the modulated power spectrum. Therefore, instead of searching the best-fit parameters, we just show one possible values of the parameters that provide better fit of the observational CMB data than the power-law primordial power spectrum. 
 
\begin{figure}[h]
		\centering
		\includegraphics[width=.7\linewidth]{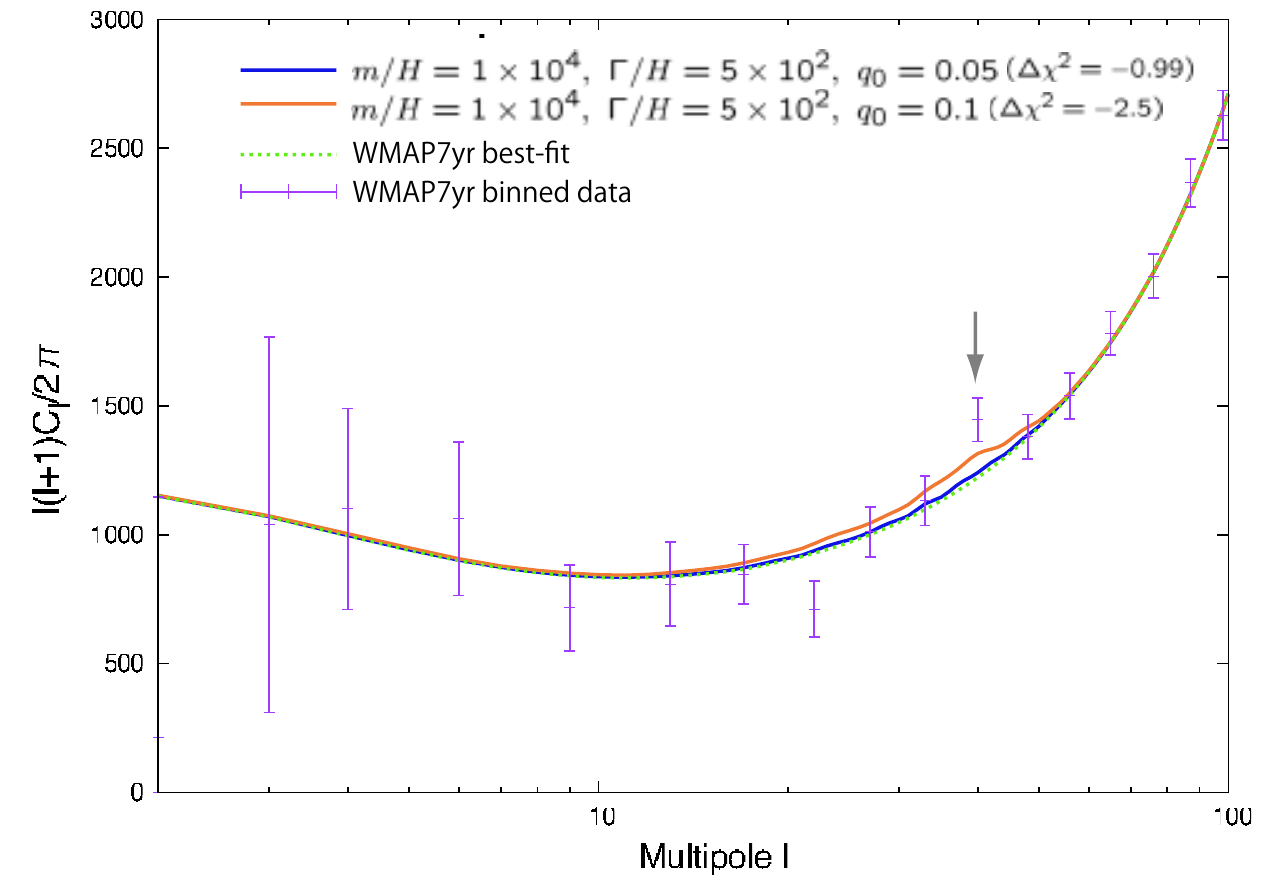}
		\caption{The angular power spectrum of the CMB for the modulated power spectrum.}
		\label{fig:cmb}
	\end{figure}

 In Fig. \ref{fig:cmb}, we have shown the angular power spectrum of the CMB for the modulated power spectrum with a feature at $k \simeq 0.003~{\rm Mpc}^{-1}$, which corresponds to multipole $l \simeq 40$. As possible values of the parameters, we have used $m/H=10^4$, $\Gamma/H=5\times10^2$, and $q_0=0.1$, while the other cosmological parameters have been kept fixed to the WMAP 7 bet-fit values \cite{Komatsu:2010fb}. Using these values of the parameters, we have found that the $\chi^2$-value impoves $2.5$ in comparison to the simple power-law spectrum. Hence, the modulated spectrum can provide a better fit of the data, though the improvement is not enough to confirm that the CMB anomalies are originated from the resonant effect of the heavy scalar field since we have introduced three additional parameters. On the other hand, for the feature at $k \simeq 0.009~{\rm Mpc}^{-1}$, it is difficult to obtain significant improvement. The angular power spectrum of the CMB is provided by a convolution of the primordial power spectrum with a radiative transfer function, which is written in terms of the spherical Bessel function. Since the radiative transfer function for multipole $l$ has an oscillatory tail for larger wavenumbers, $k\cdot(1.4\times10^4~{\rm Mpc}) \agt l$, the peak induced in the angular power spectrum has a relatively large width even if the primordial power spectrum has a sharp peak. Though the observed sharp and high peak could be realized if the modulated power spectrum has an oscillatory component coherent with that in the radiative transfer function, this is not the case for the spectra obtained here. 
   
 Though the evidence found here is not statistically significant, this will give us some insight into the possibility to find some signatures of scalar fields much heavier than the Hubble scale. If we could found such signatures, they will provide a hint for physics behind inflation.
 
\section{Summary and discussion}\label{s:summary}
 In this paper, we have discussed the possibility that a heavy scalar field, whose mass exceeds the Hubble scale, $m \gg H$, could leave non-negligible signatures in the CMB spectrum through parametric resonance between its background oscillations and the inflaton fluctuations. The resonance could be efficient without spoiling the slow-roll inflation if the heavy scalar field couples with the inflaton derivatively; the feature induced by the resonance can be sharp and large even in the case that adiabaticity is only mildly violated because the resonance coherently accumulates small effects. In the analysis, we have assumed that the oscillations of the heavy scalar field are instantaneously excited at e-folds $N_p \simeq \ln(m/H)+N_{\ast}$, where $N_{\ast}$ is the number of e-folds from the end of inflation when the peak modes crossed the horizon. This assumption will be appropriate if inflation begins at the e-folds $N_p$ and the heavy scalar field is initially displaced from its minimum, which is natural in the cases that the minima differ before and during inflation, or inflation occurs after a tunneling from a neighboring minimum \cite{Bucher:1994gb,Sasaki:1994yt,Freivogel:2005vv,Yamauchi:2011qq,Sugimura:2011tk}, for example. As another possibility, the oscillations could be dynamically excited when the heavy scalar field becomes momentarily light/tachyonic or the slow-roll condition is temporary violated. In the latter case, effects of the slow-roll violation on the resonance should be taken into account to consider large excitations of the heavy scalar field \cite{Avgoustidis:2012yc}. We will address this issue using a specific model in another paper.
  
  We have also estimated the goodness-of-fit of our model with the anomalies observed in the CMB spectrum. The resultant improvement of the fit has not been large enough to confirm that the CMB anomalies are originated from the resonance, though a systematic analysis has not been performed for technical difficulties. To test our model further from observations, non-Gaussianity in the CMB anisotropies will be helpful. Since oscillatory components are induced in the interactions, the higher-point correlation functions are also enhanced at specific scales by the resonance \cite{Chen:2008wn, Flauger:2010ja, Chen:2010bka, Behbahani:2011it}. We will estimate the amplitude and the shape of this non-Gaussianity in an upcoming paper.
 
\section*{Acknowledgement}
 The work is supported by a Grant-in-Aid through JSPS (RS, MN, YT) and was partially supported by JSPS Grant-in-Aid for Scientific Research No. 23340058 (JY) and the Grant-in-Aid for Scientific Research on Innovative Areas No. 21111006 (JY).
\appendix


\section{}\label{a:assumptions}
 Here, we examine the two assumptions made in \S \ref{s:amp}: i) the metric fluctuations can be neglected during the parametric resonance, where $k/a \sim m \gg H$ and ii) the fluctuations in the heavy scalar field, $\chi$, can be neglected.\\
 To compute the second-order action, we work in the ADM formalism,
	\begin{equation}\label{eq:admmetric}
		\mr{d}s^2 = -N^2\mr{d}t^2 + \gamma_{ij}(\mr{d}x^i + N^i\mr{d}t)(\mr{d}x^j + N^j\mr{d}t).
	\end{equation}
 Using the metric (\ref{eq:admmetric}), the action for the gravity and the scalar fields can be written as
	\begin{equation}\label{eq:admaction}
		S = \int \mr{d}t\mr{d}x^3 N\sqrt{\gamma}\left[ \frac{M_p^2}{2}R^{(3)} + \frac{M_p^2}{2}(K_{ij}K^{ij}-K^2) + P(X^{IJ}, \mb{\phi}^I) \right],
	\end{equation}
where $R^{(3)}$ is the three dimensional Ricci curvature and $K_{ij}$ is the extrinsic curvature of the time slice,
	\begin{equation}\label{eq:extrinsic}
		K_{ij} \equiv \frac{1}{2N}(\dot{\gamma}_{ij}-N_{i|j}-N_{j|i}),
	\end{equation}
Here, $|$ denotes the covariant derivative for the spatial metric $\gamma_{ij}$. We have written the terms for the scalar fields simply as $P(X^{IJ}, \mb{\phi}^I)$ because a detailed form is not important here. For the metric (\ref{eq:admmetric}), $X^{IJ}$ is written as
	\begin{equation}
		X^{IJ} = \frac{1}{2N}(\dot{\mb{\phi}}^I-N^i\pd_i \mb{\phi}^I)(\dot{\mb{\phi}}^J-N^i\pd_i \mb{\phi}^J) - \frac{1}{2}\gamma^{ij}\pd_i \mb{\phi}^I \pd_j \mb{\phi}^J.
	\end{equation}
 The Friedman equation and the evolution equations for the background fields can be written as,
	\begin{align}
		3M_p^2 H^2 = \rho, \\
		\dot{\pi}_I + 3H\pi_I - P_I = 0,
	\end{align}
where we have introduced the energy density of the scalar fields, $\rho \equiv 2P_{IJ}X^{IJ}-P$, and the conjugate momenta of the scalar fields, $\pi_I \equiv P_{IJ}\dot{\mb{\phi}}^J$.\\
 We estimate the contributions from the metric fluctuations and the fluctuations in $\chi$ by expanding the action (\ref{eq:admaction}) explicitly up to second order in the fluctuations. To eliminate the gauge degrees of freedom in the metric fluctuations, we adopt the flat gauge condition,
	\begin{equation}\label{eq:flat}
		\gamma_{ij} = a^2\delta_{ij},
	\end{equation}
where we have neglected the vector and tensor perturbations. Expanding $N$ and $N^i$ as
	\begin{align}
		N &= 1+\alpha, \\
		N_i &= \pd_i \beta,
	\end{align}
second-order terms dependent on these metric fluctuations in the action (\ref{eq:admaction}) can be written as
	\begin{equation}
	 S_{\mr{g}} = \int \mr{d}t\mr{d}x^3 a^3\Biggl[ (\rho_{IJ}X^{IJ}-\rho)\alpha^2 -\alpha\rho_{IJ}\dot{\mb{\phi}}^I\dot{\mb{\varphi}}^J - \alpha \rho_I\mb{\varphi}^I + (\pi_I\mb{\varphi}^I-2M_p^2 H\alpha)\frac{\pd^2\beta}{a^2}\Biggr], \label{eq:gaction}
	\end{equation}
where $\mb{\varphi}^I$ is the fluctuations in $\mb{\phi}^I$. This action leads to the following constraints at the first order of the fluctuations,
	\begin{align}
		\alpha &= \frac{\pi_I}{2M_p^2 H}\mb{\varphi}^I, \label{eq:alpha}\\
		\frac{\pd^2\beta}{a^2} &= \frac{\rho_{IJ}X^{IJ}-\rho}{M_p^2 H}\alpha + \frac{\rho_{IJ}}{2M_p^2 H}\dot{\mb{\phi}}^I\dot{\mb{\varphi}}^J - \frac{\rho_I}{2M_p^2 H}\mb{\varphi}^I. \label{eq:beta}
	\end{align}
Substituting Eq. (\ref{eq:alpha}) and Eq. (\ref{eq:beta}) into Eq. (\ref{eq:gaction}) the contributions from metric fluctuations can be written as
	\begin{align}
		S_{\mr{g}} &= \int \mr{d}t\mr{d}x^3 \frac{a^3}{2}\left[\hat{\mc{M}}_{IJ}^{(\mr{g})}H^2\mb{\varphi}^I\mb{\varphi}^J + \hat{\mc{N}}_{IJ}^{(\mr{g})}H\mb{\varphi}^I\dot{\mb{\varphi}}^J \right],
	\end{align}
where
	\begin{align}
		\hat{\mc{M}}_{IJ}^{(\mr{g})} &\equiv \frac{3}{2}\left(\frac{\rho_{KL}X^{KL}-\rho}{\rho}\right)\frac{\pi_I}{M_p H}\frac{\pi_J}{M_p H} - \frac{3}{2}\left(\frac{M_p\rho_{(I}}{\rho}\right)\frac{\pi_{J)}}{M_p H} + \frac{1}{2a^3 M_p^2 H^2}\dif{}{}{t}\left(\frac{a^3}{H}\pi_I \pi_J\right) \\
		&= \left[P_{MN,KL}X^{MN}\frac{\pi_I}{M_p H}\frac{\pi_J}{M_p H} - M_p P_{KL,(I|}\frac{\pi_{|J)}}{M_p H}\right]\left(\frac{X^{KL}}{M_p^2 H^2}\right) + \frac{1}{a^3 M_p^2 H^2}\dif{}{}{t}\left(\frac{a^3}{H}\pi_I \pi_J\right), \label{eq:mass}\\
		\hat{\mc{N}}_{IJ}^{(\mr{g})} &\equiv 2P_{IK,LM}X^{LM}\frac{\dot{\mb{\phi}}^K}{M_p H}\frac{\pi_{J}}{M_p H}, \label{eq:dis}
	\end{align}
which coincide with terms obtained in Refs. \cite{Langlois:2008qf, Arroja:2008yy}. Here, $P_{IJ,KL}$ and $P_{IJ,K}$ represent another derivative of $P_{IJ}$ with respect to $X^{KL}$ and $\phi^K$ respectively. These terms contribute to the equation of the motion for the fluctuations as $\hat{\mc{N}}_{[IJ]}^{(\mr{g})}H\dot{\mb{\varphi}}^J$ and $[\hat{\mc{M}}_{IJ}^{(\mr{g})}H^2 - a^{-3}(a^3\hat{\mc{N}}_{IJ}^{(\mr{g})}H)^{\cdot}/2]\mb{\varphi}^J$. At the leading order in $\eps$, $f_{\chi}$, $\mb{\phi}^{(2)}/\Lambda_n$, and $\dot{\mb{\phi}}^I/\Lambda_d^2$, $\hat{\mc{M}}_{IJ}^{(\mr{g})}$ is estimated to be
	\begin{align}
		\hat{\mc{M}}_{11}^{(\mr{g})} \sim O(\eps), \quad \hat{\mc{M}}_{12}^{(\mr{g})} \sim \frac{M_p}{\Lambda_n}O(\eps^{3/2})+\frac{m}{H}O(\eps^{1/2}f_{\chi}^{1/2}), \quad \hat{\mc{M}}_{22}^{(\mr{g})} \sim \frac{m}{H}O(f_{\chi}).
	\end{align}
On the other hand, $\hat{\mc{N}}_{IJ}^{(\mr{g})}$ is estimated to be
	\begin{align}
		\hat{\mc{N}}_{11}^{(\mr{g})} \sim \hat{\mc{N}}_{22}^{(\mr{g})} \sim \frac{\dot{\chi}^2}{\Lambda_d^4}O(\eps), \quad \hat{\mc{N}}_{12}^{(\mr{g})} \sim \frac{\dot{\phi}\dot{\chi}}{\Lambda_d^4}\left[O(\eps)+O(f_{\chi})\right],
	\end{align}
and $(\hat{\mc{N}}_{IJ}^{(\mr{g})})^{\cdot} \sim m\hat{\mc{N}}_{IJ}^{(\mr{g})}$. Hence, the contributions from the metric fluctuations are much smaller than the term $\pd^2 \mb{\varphi}^{I}/a^2$, which is estimated to be $\pd^2 \mb{\varphi}^I/a^2 \sim m^2\mb{\varphi}^I$ at $k/a \sim m$. \\ 
 On the other hand, second-order terms dependent only on the fluctuations in the scalar fields are given by
	\begin{align}
		S_{\varphi} &= \int \mr{d}t\mr{d}^3x~ \frac{a^3}{2}\left[ (2P_{IK,LJ}X^{KL}+P_{IJ})\dot{\mb{\varphi}}^I\dot{\mb{\varphi}}^J - P_{IJ}\pd_i \mb{\varphi}^I \pd^i \mb{\varphi}^J/a^2 - P_{I,J}\mb{\varphi}^{I}\mb{\varphi}^J + 2P_{KJ,I}\dot{\mb{\phi}}^K\mb{\varphi}^{I}\dot{\mb{\varphi}}^{J} \right]. \label{eq:scalaraction}
	\end{align}
 The fluctuations in the heavy scalar field, $\mb{\varphi}^{(2)}$, oscillate with a frequency larger than $m$ because of the mass term. Therefore, $\mb{\varphi}^{(2)}$ does not lead to the parametric resonance in contrast to $\mb{\varphi}^{(1)}$. Moreover, the mass term also suppresses the amplitude of $\mb{\varphi}^{(2)}$ at $k/a \sim m$ compared to that of $\mb{\varphi}^{(1)}$. Hence, we can safely discard the fluctuations in the heavy scalar field, $\mb{\varphi}^{(2)}$.\\
 Neglecting the contributions from the fluctuations in the metric and the heavy scalar field, the second-order action reduces to a rather simple form,
	\begin{align}
		S_2 &= \int \mr{d}t\mr{d}^3x~ \frac{a^3}{2}\left[ (2P_{1K,L1}X^{KL}+P_{11})\dot{\varphi}^2 - P_{11}(\nabla \varphi)^2/a^2 - P_{1,1}\varphi^2 \right]. \label{eq:infaction}
	\end{align}
 The potential term can be rewritten as $P_{1,1}\varphi^2 \simeq -3\eta_V H^2 \varphi^2$. Hence we can also neglect this term. Using the form (\ref{eq:action}) for $P$, we obtain the action (\ref{eq:2action}) in \S \ref{s:amp}.

\end{document}